\documentclass[seceq,preprint]{ptptex}

\usepackage{amsmath} \usepackage{graphicx} \usepackage{amsfonts}
\usepackage{array} \usepackage{amsthm} \usepackage{bm}\usepackage{latexsym}
\usepackage{epic} \usepackage{eepic}
\usepackage{graphicx,psfrag }

\newcommand{\la}{\lambda}
\newcommand{\lb}{\label}
\newcommand{\om}{\omega}
\newcommand{\al}{\alpha}

\newcommand{\ka}{\kappa}

\newcommand{\ga}{\gamma}

\newcommand{\bw}{\begin{widetext}}
\newcommand{\ew}{\end{widetext}}

\newcommand{\be}{\begin{equation}}
\newcommand{\ee}{\end{equation}}
\newcommand{\ba}{\begin{eqnarray}}
\newcommand{\ea}{\end{eqnarray}}
\newcommand{\non}{\nonumber}

\def\cE{{\cal E}}
\def\cG{{\cal G}}

\def\d{\partial}

\newcommand{\bei}{\begin{itemize}}
\newcommand{\eei}{\end{itemize}}

\newcommand{\e}{{\rm e}}

\title{Generating solutions via sigma-models \footnote{An
updated version of the talk given at ICGA8 and published in the
Proceedings: Prog. Theor. Phys. Suppl. {\bf 172} (2008) 121-130} }

\author{\textsc{Dmitri V. Gal'tsov}%
}

\inst{
Department of Theoretical Physics, Moscow State University, 119899,
Moscow, Russia }

\abst{
We review recent development of solution-generating techniques for
four and five-dimensional Einstein equations coupled to vector and
scalar fields. This includes D=4 Einstein-Maxwell-dilaton-axion
theory with multiple vector fields, D=5 Einstein-Maxwell gravity
with the Chern-Simons term (minimal five-dimensional supergravity),
and some other models which attracted attention in connection with
black rings. The method is based on reduction to three-dimensional
gravity coupled sigma-models with symmetric target spaces.  Our
recent results open a way to construct the general charged black
rings in five-dimensional supergravity possibly coupled to vector
multiplets.}

\begin{document}

\maketitle
\section{Introduction}
To find exact solutions of the Einstein equations is a formidable
task already in four dimensions, not to say about higher-dimensional
gravity/supergravity theories. The system is non-separable, so
usually one chooses a certain ansatz for the metric to obtain a
truncated system suitable for integration.  The choice of the ansatz
is based on the assumption that the desired solutions possess some
isometry group.   Assumption of the symmetry means that all the
unknown functions depend on a number of variables smaller than the
initial dimension $D$ of the space-time.  This leads to dimensional
reduction of Einstein equations to a lower-dimensional system
involving additional vector and scalar fields. An especially
fruitful technique is based on the assumption of an Abelian isometry
group of the rank $D-3$. This gives rise to the toroidal reduction
to three dimensions, where vector fields (and higher rank
antisymmetric forms, if present) can be dualized to scalars, and one
obtains three-dimensional gravity coupled to the set of purely
scalar fields. The remarkable feature of vacuum gravity in any
dimension as well as the bosonic sectors of ungauged supergravities
is that these scalar fields form sigma-models on coset spaces. The
residual of the diffeomorphism group acting in the reduced
dimensions together with symmetries of the vector multiplets combine
into an enhanced duality group of global symmetries, often termed as
``hidden symmetries''. The existence of hidden symmetries gives rise
to a variety of tools like generation of new solutions from known
ones, construction the BPS solutions as null geodesics of the target
spaces, and derivation of two-dimensional integrable systems.

Hidden symmetry $SL(2,R)$ of the four-dimensional vacuum Einstein
gravity reduced to three dimensions was discovered by Ehlers
\cite{eh}, the corresponding two-dimensional integrable systems were
explored in a number of papers \cite{ern,int} (see for a review
\cite{ex2,vd,bbook}). The group $SL(2,R)$ is always present as a
part of hidden symmetries of more complicated systems,  so it
remains an ingredient of generating techniques for them too. The
next in complexity is the four-dimensional Einstein-Maxwell (EM)
theory (or bosonic part of $D=4,\,N=2 $ supergravity). A novel
feature is presence of the Harrison transformations  which generate
charged solutions from neutral ones\cite{nk,ki,har,mz,ggk}~. Vacuum
gravity in five and higher dimensions was studied along the same
lines in\cite{ne,ms}~. Later it was realized that various
higher-dimensional supergravities also lead to the three-dimensional
gravity coupled sigma-models with symmetric target spaces
\cite{ju,julia81,bgm,sen2}~. A very convenient dimensional reduction
scheme was elaborated in \cite{cjlp1} (see also \cite{cjlp}) for the
toroidal reduction of maximal supergravities originating from
eleven-dimensional supergravity. It allows to obtain the Cartan-Weyl
root system of the hidden symmetry group directly in terms of the
so-called dilaton vectors --- the coefficients of the dilatonic
exponents emerging in the dimensional reduction.

Three-dimensional sigma models resulting from  supergravities are
quite complicated generically, so the possibility of their
application for solution generation is rather limited. Meanwhile
some physically interesting truncated systems  other than vacuum and
$D=4$ electrovacuum were discovered which are relatively simple and
lead to manageable models. One such system is the $D=4$
Einstein-Maxwell-dilaton-axion (EMDA) gravity which is a truncated
version of the $N=4$ supergravity (or toroidally compactified
heterotic string theory) \cite{gk,g,diak,cg}~. It contains one
Maxwell field and two massless scalar fields (dilaton and axion) and
leads in three dimensions to the sigma-model with the global
symmetry $SO(2,3)\sim Sp(4,R)$. It admits a representation in terms
of $4\times 4$ matrices which can be  split into the $2\times 2 $
blocks. It also allows for K\"{a}hler representation in complex
variables similar to Ernst potentials in vacuum and electrovacuum
$D=4$ gravity. This representation can be generalized to the EMDA
gravity with an arbitrary number of vector fields \cite{gl}~.
Especially interesting is the case of two vector fields  which gives
rise to a quaternion analog of the Ernst potential \cite{gsh}~.

Recent interest to black rings (for a review see \cite{brin})
stimulated investigation of hidden symmetries of five-dimensional
gravity coupled to vector fields. Somewhat unexpectedly,
five-dimensional Einstein-Maxwell gravity does not possess
non-trivial hidden symmetries  which can generate charges (the
isometry group of the target space of the corresponding
three-dimensional sigma-model is solvable). This explains why no
analytic solutions are known for charged rotating black holes in
this theory. However there exist more complicated five-dimensional
gravity theories with vector fields which possess  non-trivial
hidden symmetries containing charging (Harrison) transformations.
One of them corresponds to dimensional reduction of six-dimensional
vacuum gravity to five dimensions. It contains a Kaluza-Klein vector
field and a dilaton, the hidden symmetry being $SL(4,R)$. A matrix
representation for this theory similar to that for EMDA was
constructed in \cite{chgsh}~. The second is five-dimensional minimal
supergravity whose bosonic sector is the Einstein-Maxwell theory
with the Chern-Simons term \cite{cre,CN}~. The presence of the
latter leads to a non-trivial hidden symmetry realized by an
exceptional group $G_{2(2)}$ (the non-compact version of $G_2$)
\cite{mizo,PS}~. A detailed study of the corresponding
three-dimensional sigma-model was performed in \cite{six} (see also
\cite{gc07}). A more general five-dimensional model containing three
vector and two independent scalar fields arising in some truncated
toroidal reduction of $D=11$ supergravity is often invoked in the
discussion of black rings \cite{brin}~. As we have shown recently
\cite{gshch08}, the hidden symmetry in this case is $SO(4,4)$.
\section{D=4}
\subsection{Ehlers group  SL(2,R)\cite{eh}}
Consider vacuum Einstein equations $R_{\mu\nu}=0$. For stationary
metrics admitting a time-like Killing vector field,  $ {\cal L}_K
g_{\mu\nu}=0,\;  K=\partial_t$,  the line element can be presented
as \vspace{-.1cm}
\begin{equation} ds^2=-\e^\xi(dt+\omega_i dx^i)^2+\e^{-\xi}h_{ij}dx^idx^j,
\end{equation}
where the scalar $\xi$, the three-vector $\omega_i$ and the
three-dimensional metric $h_{ij}$ depend only on spatial coordinates
$x^i$. Then the equations of motion  coincide with those of the
three-dimensional gravitating non-linear sigma-model \vspace{-.1cm}
\begin{equation}  S_\sigma=\int\left[R_3(h)-\cG_{AB}(\Phi)\d_i\Phi^A\d_j\Phi^B  h^{ij}\right]
\sqrt{h}d^3x,  \vspace{-.1cm}  \end{equation} with two scalar fields
$\Phi^A =(\xi,\,\chi)$,  where the twist potential  $\chi$ is
related to the one-form $\omega_i$ by the dualization equation
$d\chi=-\e^{2\xi}*d\om$, and the target space is a coset space
  $SL(2,R)/SO(1,1)$, the corresponding  metric being \vspace{-.1cm}
\begin{equation}
dl^2=\cG_{AB}d\Phi^A\d\Phi^B=\frac12\left(d\xi^2+\e^{-2\xi}d\chi^2\right).
 \vspace{-.1cm} \end{equation} Now the initial four-metric
$g_{\mu\nu}$ is presented by the three-metric $h_{ij}$ and two
"matter" fields  $\xi,\,\phi$, the field equations being invariant
under  the Ehlers group $SL(2,R)$ acting transitively on the target
space as an isometry: $\xi,\,\chi\to\xi',\,\chi'$. So starting with
some solution $g_{\mu\nu}=\left\{h_{ij}, \xi', \om'\right\}$
 one can find another set $\xi',\,\chi'$.  Dualizing back
$\chi'$ to $\om_i'$ one obtains a new 4D solution with the same
three-metric
 $g_{\mu\nu}=\left\{h_{ij}, \xi', \om'\right\}$.
Three $SL(2,R)$ transformations consist of \par $i)$ twist shift
(gauge) $\chi \to\chi+\la_g$,\par $ii)$ scaling $\xi \to\xi+\la_s
,\;\;\chi\to\e^{\la_s}\chi$,\par $iii)$ proper Ehlers transformation
$(\chi-i\e^\xi)^{-1} \to (\chi-i\e^\xi)^{-1} +\la_E$,
\\ the last one generating the Taub-NUT metric from Schwarzschild.

Similarly, vacuum gravity in D dimensions has hidden symmetry
$SL(D-2,R)$ \cite{ms}~.
\subsection{Einstein-Maxwell}
The sigma-model representation can be derived for 4D gravity coupled
to a massless vector field $S=-\frac{1}{16\pi}\int \left(R +
 F^2\right)\sqrt{-g}\;d^4x.$   With the assumption of stationarity,
the $\mu =i$ component of the Bianchi identity
 $\partial_\nu(\sqrt{-g}{\tilde F}^{\mu\nu})=0$  is satisfied
introducing the electric potential  $ F_{i0}= \partial_iv,$
  while the  $\mu =i$ component of the Maxwell
equation  $\partial_\nu(\sqrt{-g} F^{\mu\nu})=0$   is solved by
introducing the magnetic potential  $u$:
$F^{ij}=\frac{\e^\xi}{\sqrt{ h}} \epsilon^{ijk}\partial_ku. $  The
remaining components of $F^{\mu\nu}$ can be expressed in terms of
$v$ and $u$, so a vector field is reduced to  two scalars. The
resulting 3D sigma-model is especially simple in the static case
 $\om_i=0=\chi$. Then  the full
Einstein-Maxwell system splits into independent electric and
magnetic sectors. In the electric case one obtains $SL(2,R)/SO(2)$
sigma-model with the target metric
 $dl^2=
\frac12 d\xi^2-\e^{-\xi}dv^2$,  in the magnetic case $dl^2= \frac12
d\xi^2-\e^{-\xi}du^2. $ Thus we deal again with the $SL(2,R)$
transformations acting on $\xi, v$ or $\xi, u$. The last one,
$iii)$,  now is the Harrison transformation generating charged
solutions from uncharged ones. Taking the Schwarzschild solution as
a seed, one can construct the Reissner-Nordstr\"om solution.

 For a general stationary EM system one gets the four- dimensional
target space with the signature $(++--)$:
\vspace{-.1cm}\begin{equation}
dl^2=\frac12\left[d\xi^2+\e^{-2\xi}(d\chi+vdu-udv)^2\right]-\e^{-\xi}(dv^2+du^2),
\vspace{-.1cm}\end{equation}  which is the coset space
$SU(2,1)/S(U(1)\times U(1))$ \cite{ki}~. Complex Ernst potentials
\cite{ern,ki}
 $ \Phi=\frac1{\sqrt2}(v+iu),\;
\cE=\e^\xi+i\chi+\Phi\Phi^*  $ realize a non-linear representation
of $SU(2,1)$, whose action consists of three gauge (shifts of $\chi,
v, u$), an electric-magnetic rotation, a scaling, two charging
Harrison transformations  \vspace{-.1cm} \begin{equation}
\Phi\to\frac{\Phi+c\cE}{1-2c^*\Phi-|c|^2\cE},\;\;
\cE\to\frac{\cE}{1-2c^*\Phi-|c|^2\cE},\vspace{-.1cm}
\end{equation} and Ehlers transformation $\Phi\to \Phi
(1+i\ga\cE)^{-1},\;\cE\to \cE (1+i\ga\cE)^{-1}$ where $c$ is complex
and $\ga$ is real parameters.
\subsection{EM-Dilaton}   Adding a dilaton with an arbitrary coupling
 \vspace{-.1cm} \begin{equation}S=\frac{1}{16\pi}\int
\left(-R+2(\partial\phi )^2-
\e^{-2\alpha\phi}F^2\right)\sqrt{-g}\;d^4x,\vspace{-.1cm}
\end{equation} one obtains the target space
 \vspace{-.1cm}\begin{equation} dl^2 \!\! = \!\!  \frac12 \left[d\xi^2\!\! + \! \e^{-2\xi}
(d\chi\!\! +\!  vdu \!\!-\!  u dv)^2 \!\right]\!+ \!2d\phi^2\!\!- \!
\e^{-\xi} (e^{-2\alpha\phi}dv^2 \!\!+
\!e^{2\alpha\phi}du^2).\vspace{-.1cm}
\end{equation} It is a symmetric space only for $\al^2=0,\,3$
\cite{ggk}~. In the first case we have EM + a decoupled scalar, in
the second -- a compactified 5D vacuum gravity (Kaluza-Klein), in
which case the isometry group is 8-parametric $SL(3,R)$. For other
$\al$ the isometry group is a 5-parametric solvable group not
containing Harrison-Ehlers transformations.

It is worth noting, that contrary to vacuum gravity, the
Einstein-Maxwell theory in $D>4$ does not possess semi-simple hidden
symmetries.
\subsection{EM-Dilaton-Axion}  The previous model has only
a discrete electric-magnetic duality. Adding an axion $\ka$ we
obtain the EMDA model (a truncated $N=4$ supergravity) which possess
a continuous $SL(2,R)$ electric-magnetic symmetry
\begin{equation} \label{an}
S=\frac{1}{16\pi}\int \left\{-R+2\partial_\mu\phi\partial^\mu\phi +
\frac{1}{2} e^{4\phi} {\partial_\mu}\kappa\partial^\mu\kappa
-e^{-2\phi}F_{\mu\nu}F^{\mu\nu}-\kappa F_{\mu\nu}{\tilde
F}^{\mu\nu}\right\} \sqrt{-g}\,d^4x,
\end{equation}
where ${\tilde
F}^{\mu\nu}=\frac{1}{2}E^{\mu\nu\lambda\tau}F_{\lambda\tau},\;
F=dA\;$. The resulting target space is the 6-dimensional coset
$Sp(4,R)/(SO(2)\times SO(1, 2))$ \cite{gk,g,diak,cg}:
\begin{equation} \lb{emdat} \!\!dl^2 \!\!= \frac12 \left[d\xi^2\!\!
+ \! \e^{-2\xi}(d\chi+vdu-udv)^2+ \!\!\e^{4\phi} d\kappa^2\right] +
2d\phi^2 -\e^{-\xi}\left[e^{-2\phi}dv^2\!\!+e^{2\phi}(du-\kappa
dv)^2\right]\!\!.  \end{equation}
 Its isometry group $Sp(4,R)\sim SO(3,2)$
consists of three gauge, one scale, three SL(2,R) S-duality, two
Harrison and Ehlers transformations (altogether 10). All isometry
transformations are known algebraically.
\section{Matrix methods}
 To find symmetry
transformations explicitly one can  use an appropriate matrix
representative
 $ M=M(\Phi)$ of the coset, so that the target space metric is presented as
 \vspace{-.1cm} \begin{equation} dl^2=\cG_{AB}\,d\Phi^Ad\Phi^B=
-k{\rm Tr}\left(dMdM^{-1}\right). \vspace{-.1cm}\end{equation}
 One-parametric subgroups  $G_\la=\e^{\la \bar K}$
corresponding to the Killing  vector $K$  ($\bar K$ being the
relevant matrix generator) act on $M $  as  $M'= G_\la^{-1}MG_\la$ .
Then the transformation of the potentials can be read off from the
relation
 $M'(\Phi)=M(\Phi').$
 For EMDA, $M $  is
given by a symmetric symplectic  $4\times 4$ matrix \cite{diak}:
\begin{equation} M=\left(\begin{array}{crc}
P^{-1}&P^{-1}Q\\
QP^{-1}&P+QP^{-1}Q\\
\end{array}\right),\end{equation}
  where $P$ and $Q$ are the real symmetric $2 \times 2$
matrices
\begin{equation} \!P=-{\rm e}^{-2\phi} \left(\begin{array}{crc}
\!\!v^2-{\rm
e}^{2\phi+\xi}&\!\!\!\!v\\
\!\!v&1\\
\end{array}\!\!\right),\;\;Q=\left(\begin{array}{crc}
vw-\chi&w\\
w&-\kappa\\
\end{array}\right),  \end{equation} $w=u-\ka v$. Further simplification consists in
presenting ten Sp(4,R) isometries as SL(2,R) transformation acting
on 2x2 complex matrices   $Z=P+iQ$ \vspace{-.1cm}
\begin{equation} Z=\left(\begin{array}{crc}
E&\Phi\\
\Phi&-z\\
\end{array}\right),\quad \begin{array}{crc}
z=\ka+i\e^{-2\phi}, & \Phi=u-zv, \\
 E=i\e^{\xi}-\chi+v\Phi,&\\
\end{array} \vspace{-.1cm} \end{equation}
where the quantities $\Phi,\, E$ can be viewed as the EMDA
Ernst-type complex potentials. Isometry transformations consist of
shift  and  scaling  of $Z$, and  shift of $1/Z$: \vspace{-.1cm}
\begin{equation}\lb{z1}
  Z\to Z+B,\;\;  B=\left(\begin{array}{crc}
g&m\\
m&-b\\
\end{array}\right),\;\; Z\to A^T Z A,\;\; A=\left(\begin{array}{crc}
{\rm e}^s & h_e\\
-e  & \e^a \\
\end{array}\right), \vspace{-.1cm}
\end{equation}  \begin{equation}\lb{z2}
\frac1Z\to  \frac1Z+C,\;\;  C=\left(\begin{array}{crc}
c_E & h_m\\
h_m & -c\\
\end{array}\right).  \vspace{-.1cm} \end{equation}
Real parameters correspond to: $g, s$  -- gravitational gauge, $e,
m$ -- electric and  magnetic gauge;  $a,b,c$ -- $SL(2,R)$ S-duality,
$h_e,\; h_m$  -- electric and magnetic Harrison transformations,   $
c_E$ -- Ehlers transformation. $SL(2,R)$ S-duality was the symmetry
of the 4D theory, acting on the axidilaton $z$ and the Maxwell
field, now it is incorporated into the 3D U-duality group $Sp(4,R)$.

Further simplification comes from the fact that the target space
(\ref{emdat}) admits K\"{a}hler complex parametrization $z_\alpha,\,
\alpha=0,1,2$ with $z_0=E-z,\;z_1=u-v\Phi,\; z_2=E+z$. The
K\"{a}hler target space metric is constructed from the K\"{a}hler
potential:\begin{equation}  \cG_{\alpha {\bar
\beta}}=\partial_\alpha
\partial_{\bar\beta} K(z^\alpha, {\bar z}^\beta), \quad K=-\ln V,\quad
V= \eta_{\alpha\beta}{\rm Im} z^\alpha {\rm Im} z^\beta = {\rm
e}^{\xi-2\phi}, \end{equation} where  $\eta_{\alpha\beta}={\rm diag}
(-1,1,1). $  This construction can  be directly generalized to the
case of an arbitrary number $p$ of vector fields  introducing
additional complex coordinates $z_n = u_n-z v_n\equiv \Phi_n ,\;
n=1,\ldots,p$. Remarkably, the K\"{a}hler potential  remains the
same $V={\rm e}^{\,\xi-2\phi}$ being expressed in terms of the real
quantities \cite{gl}, though its realization in complex variables
now involves the metric
 $\eta_{\alpha\beta}={\rm diag} (-1,1,...,1),\,\alpha,
\beta=0,1,...,p+1$. The target space of EMDA with $p$ vector fields
is  the coset   $SO(2,2+p)/\left(SO(2)\times SO(p,2)\right)$, its
matrix representation is realized by $(p+4)\times(p+4)$ matrices.

The case $p=2$ is exceptional: due to isomorphism   $SO(2,4)\sim
SU(2,2)$, the $4\times 4$ representation is possible instead of the
expected $6\times 6$. It is generated by the  $2\times 2$ complex
matrix $Z$ in the same way as before, but now with a generic $Z$
incorporating four independent complex potentials:
\cite{gsh}\vspace{-.1cm}
\begin{equation}Z=\left(\begin{array}{crc}
E & \Phi_1-i\Phi_2\\
\Phi_1+i\Phi_2 & -z\\
\end{array}\right).\vspace{-.1cm}\end{equation}
Moreover, the target space metric can be directly expressed through
$Z$ as follows \vspace{-.1cm}\begin{equation} dl^2=-2{\rm
Tr}\left\{d Z \left({Z}^{\dagger}- {Z}\right)^{-1}
d{Z}^{\dagger}\left(
{Z}^{\dagger}-{Z}\right)^{-1}\right\},\vspace{-.1cm}\end{equation}
while the symmetry transformations  are again (\ref{z1}, \ref{z2}).
This realizes   the  quaternion $SL(2,Q)$  representation
 of $SU(2,2)$. Complex quaternion coordinates can be read from an
 expansion of $Z$ in terms of Pauli matrices $Z=z^0
 I_2+z^a\sigma_a$.
\section{D=5}\vspace{-.1cm}
\subsection{Black rings}
  Recent interest to  exact solutions in
five-dimensional gravity is related to possibility of topologically
non-spherical stationary asymptotically flat black holes, with the
$S^1\times S^2$ topology of the event horizon \cite{brin}~. While in
4D the uniqueness theorems imply that the most general stationary
black hole is Kerr for vacuum and Kerr-Newman for electrovacuum
(both with $S^2$ horizon topology), in 5D it is not so. It is
possible that black holes with the horizon topology $S^1\times
S^{D-3}$ exist for any D. The most general non-singular 5D black
ring is a three-parametric solution endowed with a mass and two
independent rotation parameters \cite{pse}~. The most general
charged black ring in 5D gravity coupled to a single Maxwell and
possibly scalar fields should be five-parametric, with an  electric
charge and a magnetic dipole moment as additional parameters. Such
solution is still unknown. Another interesting model includes three
vector and two independent scalar fields, within this theory a
nine-parametric solution should exist (also unknown).

This stimulates further investigation of 5D theories which lead to
3D sigma-models on symmetric spaces. Somewhat unexpectedly, pure
Einstein-Maxwell theory in 5D generates only trivial (solvable)
hidden symmetry algebra. Two other theories which possess
non-trivial hidden symmetries  are discussed below.

\subsection{5D EM-dilaton (Kaluza-Klein)}
Compactifying  6D vacuum gravity on a circle, one obtains 5D
Einstein-Maxwell-dilaton theory with the Kaluza-Klein dilaton
coupling $\alpha^2=8/3$: \vspace{-.1cm} \be S_5 = \int d^5x
\sqrt{-g_5} \left\{ R_5
    - \frac12 (\partial { \phi})^2
    - \frac{e^{-\alpha \phi}}{12} {  H}^2 \right\}, \vspace{-.1cm}\ee
    where $ \phi$ is the dilaton and
${  H}=d{  B}$ is an antisymmetric three-form dual to the KK vector
in 5D. This model obviously have the $SL(4,R)$ hidden symmetry in
3D. It is instructive to present it in a form similar to that of 4D
EMDA \cite{chgsh}~. In fact, in 4D this action contains two vector
and three scalar fields, differing from EMDA with two vector fields
by presence of an additional scalar $\psi$, as well as by different
interaction structure. The target space of the corresponding 3D
sigma-model has the line element \vspace{-.1cm} \ba \!\!\!
\!\!\!\!\!\! \!\!\! dl^2 = \frac12 \left( d\xi^2 + \e^{-2\xi}\left[
d\chi
  + \frac12 \left( v_a du_a - u_a dv_a \right) \right]^2 \right)
+\frac12 d\phi^2 + \frac14 d\psi^2
  + \frac12 \e^{2\phi} d\kappa^2 && \non\\
-\frac{\e^{-\xi}}2 \Big[ \e^{\psi-\phi} dv_1^2
  + \e^{-\psi+\phi} (du_1 - \kappa dv_2)^2
 + \e^{-\psi-\phi} (dv_2)^2 + \e^{\psi+\phi} (du_2 - \kappa dv_1)^2
    \Big],\quad&&\vspace{-.1cm}\ea where $a=1,2$.
This is the metric of the symmetric space $SL(4,R)/SO(2,2)$. As the
coset representatives one can choose the symmetric $SL(4,R)$ matrix
\vspace{-.1cm}\be M =\left( \begin{array}{cc}
   P_1^{-1} & P_1^{-1} Q \\ Q^T P_1^{-1} & P_2 + Q^T P_1^{-1} Q
   \end{array} \right),\vspace{-.1cm}
\ee where   $Q$ is  a real $2 \times 2$ matrix and $P_1, P_2$ are
symmetric matrices with the same determinant. This matrix can be
regarded as a  generalization of the EMDA $Sp(4,R)$ matrix to which
it reduces for $P_1=P_2$ and $Q^T=Q$. Now we have \vspace{-.1cm}\be
P_1 =e^{\psi/2} \left( \begin{array}{cc}
       e^{\xi-\psi} - (v_1)^2 e^{-\phi} & -v_1 e^{-\phi} \\
      -v_1 e^{-\phi} & -e^{-\phi}
      \end{array} \right), $$
 $$P_2 = e^{-\psi/2} \left( \begin{array}{cc}
      e^{\xi+\psi} - (v_2)^2 e^{-\phi} & -v_2 e^{-\phi} \\
      -v_2 e^{-\phi} & -e^{-\phi}
      \end{array} \right),\;\;
Q =\left( \begin{array}{cc}
      \frac1{2} \mu-\chi & u_2 - \kappa v_1 \\
      u_1 - \kappa v_2 & -\kappa
      \end{array} \right),\vspace{-.1cm}
 \ee where $ \mu = v_1 \left( u_1 - \kappa v_2 \right)
    + v_2 \left( u_2 - \kappa v_1 \right).$
Transformations ${  M} \to {\cal G}^T {  M} {\cal G}$ with constant
${\cal G}\in SL(4,R)$ can be presented by its action on 2x2 matrices
$P_1,\,P_2,\,Q$ similarly to the above quaternion form of $SU(2,2)$
\cite{chgsh}~.
\subsection{5D minimal supergravity}
Another symmetric model is minimal 5D supergravity containing in
addition to the Maxwell term also the Chern-Simons term. This
enlarges the hidden symmetry of the theory, the 3D U-duality being
the non-compact form of the exceptional group $G_2$.
Five-dimensional minimal supergravity contains a graviton, two
symplectic-Majorana gravitini (equivalent to a single Dirac
gravitino), and a $U(1)$ gauge field. The bosonic part of the
Lagrangian is very similar to that of $D=11$ supergravity:
\vspace{-.1cm}\be S_5 = \frac1{16 \pi G_5} \left[ \int d^5x \sqrt{-
  g} \left(   R - \frac14   F^2 \right) - \frac1{3 \sqrt3}
\int  F \wedge   F \wedge   A \right], \vspace{-.1cm} \ee where $  F
= d A$.  This action can be obtained  as a suitably truncated
Calabi-Yau compactification of $D=11$ supergravity. Its reduction to
3D leads to the sigma model on a symmetric space
$G_{2(2)}/(SL(2,R)\times SL(2,R))$ \cite{mizo,PS}~. The
corresponding generating technique was developed in \cite{six}. The
target space metric reads ($v_1,u_1$ being KK, $v_2,u_2$ -- Maxwell
potentials)\vspace{-.2cm} $$ dl^2 \!=\!
 \frac12 \!\Bigl\{\! d\xi^2 + \mathrm{e}^{-2\xi}\!\left(d\chi\! + \!v_1 d u_1\! +
\!v_2 d u_2 \!\right)^2 + 3 \left( d\phi^2 \!+\! \mathrm{e}^{2\phi}
d\kappa^2 \right)-  \mathrm{e}^{-\xi}\Bigl[\mathrm{e}^{-\phi} \left(
d v_2 \!+ \!\sqrt3   \kappa d v_1\! \right)^2
 $$\vspace{-.3cm} \be+ \mathrm{e}^{-3\phi} d v_1^2 +
\mathrm{e}^{3\phi} \left( d u_1 \!+ \!\kappa^3 d v_1 \!- \!  \sqrt3
\kappa (d u_2\! -\! \kappa d v_2) \right)^2
  +  \mathrm{e}^\phi \left( d u_2 \!-\! 2 \kappa d v_2 \!-\!
 \sqrt3  \kappa^2 d v_1 \right)^2 \Bigr]\Bigr\}~.
\vspace{-.2cm}\ee This is the metric on the coset
$G_{2(2)}/(SL(2,R)\times SL(2,R))$. Note that the number of
variables is 8, the same as for 4D EMDA with two vector fields.
However, the isometry group is 14-parametric (15 parametric for
EMDA). The matrix representation can be given either in terms of
$7\times 7$ or $8\times 8$  matrices corresponding to embedding  of
$G_{2(2)}$ into $SO(3,4)$ or $SO(4,4)$. In the first case
\vspace{-.1cm}\be M = \left(\begin{array}{ccc}
A & B & \sqrt2U \\
B^T & C & \sqrt2V \\
\sqrt2U^T & \sqrt2V^T & S
\end{array}\right),\vspace{-.1cm}\ee
where $A,B,C$ are $3\times 3$ matrices and $U,V$ are 3-columns,
whose explicit form is given in \cite{six}~. This matrix realizes a
noncompact coset  $G_{2(2)}/(SL(2,R)\times SL(2,R)).$ In spite of
being rather complicated, this representation opens a way to
construct new solutions acting on this matrix by one-parametric
subgroups of $G_{2(2)}$ which are easily found by exponentiating the
generators. The fourteen transformations include five gauge, one
scale, three S-duality, four Harrison and Ehlers transformations. A
charged rotating 5D black hole with two rotation parameters was
generated in \cite{six} from the solution \cite{pse} as a seed. It
is plagued with a conical singularity. To get a regular charged
doubly rotating solution one should start with a more general doubly
rotating vacuum ring solution \cite{mty} with non-compensated
conical singularity.
\subsection{5D SUGRA with vector multiplets}
More general models of this type containing multiple $U(1)$ vector
fields give rise to black rings with many electric and dipole
charges. One popular model leading to three-charge black rings
contains three vector and three scalar fields subject to a
constraint:\vspace{-.3cm} \be\label{L5} \!\!\!\!\!S  = \!\frac{1}{16
\pi G_5}\! \int \!\left( R_5 \star  1 - \frac12 G_{IJ}\! \left(dX^I
\wedge \star dX^J - F^I \wedge \star  F^J\right) - \frac{
\delta_{IJK}}6 F^I \wedge F^J \wedge A^K \right),\vspace{-.2cm}\ee
where $G_{IJ}={\rm diag}\left((X^1)^{-2},\ (X^2)^{-2},\
(X^3)^{-2}\right),\quad F_{I}=dA_{I},\quad I,J,K=1,2,3, $ and
$\delta_{IJK}=1$, if $ I,J,K $ is a permutation of 1,2,3, and zero
otherwise. 3D reduction of this theory leads to the sigma model on
the homogeneous space $SO(4,4)/SO(4)\times SO(4)$ or
$SO(4,4)/SO(2,2)\times SO(2,2)$ depending on the signature of the
three-space. A $8\times 8$ matrix representation  which can be used
for solution generating purposes was constructed \cite{gshch08}~. It
opens a way to find the nine-parametric charged black ring in five
dimensions. An identification of the three vector fields $A^I$ with
the corresponding contraction of the scalars $X^I$ returns us to the
$G_{2(2)}/(SL(2,R)\times SL(2,R))$ sigma model of the previous
section.
\section{Conclusions}
We have reviewed the main ideas of solution generating techniques
based on dimensional reduction to three-dimensions and presented
some new sigma-models which can be used to construct charged black
holes and black rings in five dimensions. These models can be
further reduced to two dimensions giving new integrable systems
based on $G_{2(2)}$ and $SO(4,4)$ groups.

\vspace{-.1cm}
\section*{Acknowledgements}
We would like to thank Yukawa Institute for Theoretical Physics and
Nara Women's University for hospitality and support during the ICGA8
and the GC workshop. Useful discussions with M. Sasaki and K.-I.
Maeda, R. Kallosh, Y.M. Cho and S. Odintsov are gratefully
acknowledged. We  thank G. Clement, C.-M. Chen and N. Scherbluck for
collaboration. The work was supported by RFBR under the project
08-02-01398.

%

\end{document}